# Modulated Interlayer Exciton Properties in a Two-Dimensional Moiré Crystal

Xiaobo Lu[1], Xiaoqin Li[2, 3], and Li Yang[1, 4]

[1]Department of Physics, Washington University in St. Louis, St. Louis, MO 63130, USA
[2]Department of Physics and Center for Complex Quantum Systems, University of Texas at Austin, Austin, TX 78712, USA.
[3]Texas Materials Institute, University of Texas at Austin, Austin, TX 78712, USA.
[4]Institute of Materials Science and Engineering, Washington University in St. Louis, St. Louis, MO 63130, USA

## Abstract

Twisted van der Waals heterostructures and the corresponding superlattices, moiré superlattices, are remarkable new material platforms, in which electron interactions and excited-state properties can be engineered. Particularly, the band offsets between adjacent layers can separate excited electrons and holes, forming interlayer excitons that exhibit unique optical properties. In this work, we employ the first-principles GW-Bethe-Salpeter Equation (BSE) method to calculate quasiparticle band gaps, interlayer excitons, and their modulated excited-state properties in twisted $MoSe_2/WSe_2$ bilayers that are of broad interest currently. In addition to achieving good agreements with the measured interlayer exciton energies, we predict a more than 100-meV lateral quantum confinement on quasiparticle energies and interlayer exciton energies, guiding the effort on searching for localized quantum emitters and simulating the Hubbard model in two-dimensional twisted structures. Moreover, we find that the optical dipole oscillator strength and radiative lifetime of interlayer excitons are modulated by a few orders of magnitude across moiré supercells, highlighting the potential of using moiré crystals to engineer exciton properties for optoelectronic applications.

## I. Introduction

Recently, twisted bilayers of two-dimensional (2D) van der Waals (vdW) structures have ignited significant interests. For example, twisted bilayer graphene (tBLG) [1] exhibits Mott insulating behaviors [2] and unconventional superconductivity at magic angles [3]. Many interesting properties are also expected in twisted bilayers of semiconductors, such as transition metal dichalcogenides (TMDs). Of particular interests are TMD heterostructures with a twisting angle, in which moiré superlattices with a relatively large period may form. Such moiré superlattices may realize a regular array of quantum dots defined by the in-plane potential variations formed in moiré superlattices. Topological insulators, mosaic patterns, novel spin selection rules, optical dichroism, and Hubbard models for correlated electrons are predicted through this periodical modulation as well. [4–10] Meanwhile, recent experiments have shown that the influence of moiré potentials on exciton resonances in twisted TMD bilayers may be observed even in far-field optical spectroscopy studies. [11–15]

Most of these predictions and measurements are associated with many-body physics and excited-state properties. It is known that excitonic (electron-hole ($e$-$h$)) effects are substantially enhanced in 2D semiconductors and dominate quasiparticle and optical properties. [16–19] To calculate excitonic effects, one must go beyond the ground-state density functional they (DFT) and include many-electron interactions more accurately. To date, many-body perturbation theory (MBPT) has been proven to be a well-established excited-state theory to calculate quasiparticle energies and excitons in solids. [20–22] In fact, there have been numerous MBPT calculations on TMD heterostructures, [23–26] which revealed enhanced excitonic effects and approaches to engineer $e$-$h$ pairs and their optical responses. On the other hand, how excited quasiparticles and excitons are spatially modulated in the presence of a moiré superlattice remains unknown, and an *ab initio* calculation is crucial for understanding and predicting these excited states.

In this work, we focus on interlayer exciton properties of MoSe$_2$/WSe$_2$ heterostructures with a small twist angle. [11,12,27–29] First-principles calculations of bilayer structures with small twisting angles are known to be challenging [20,30,31] because of the large supercell size. Encouraged by previous studies of interpolations of local electronic structures of twisted

TMDs, [5,6] we will focus on specific stacking styles, calculate the local quasiparticle and exciton properties, and obtain the overall spatially modulated properties by interpolation between these local sites. Given that the sizes of typical excitons of 2D TMDs are usually around a few nm, [16,19,32] the above interpolation scheme is appropriate for TMD bilayers with a small-twist angle bilayers (up to ~ 2 degree) [7,33], whose moiré period is more than 10 nm. More recently, it was reported that the shear solitons at stacking domain boundaries influence the domain size of different stacking styles. [30] This structure optimization around domain boundaries may impact the spatial extension of different stacking styles but the fundamental picture of the above interpolation shall still be valid.

The article is organized in the following order. In section II, we introduce two types of twisted $MoSe_2/WSe_2$ bilayers as well as our simulation setups. In section III, the calculations of quasiparticle energies are presented for different local stacking styles. In section IV, we present the optical absorption spectra, excitonic effects, and optical oscillator strength of interlayer excitons. In section V, we calculated the radiative lifetime of interlayer excitons. In section VI, we include higher-order impacts, i.e., the strain effect and spin-orbit coupling (SOC), to correct interlayer exciton energy and compare with experiments. The moiré patterns of interpolated interlayer excitons are presented as well. Finally, the conclusion is summarized in section VII.

**II. First-principles simulation setups and atomic structures**

We employ DFT to study the ground-state properties of $MoSe_2/WSe_2$ heterostructures. The relaxed atomistic structures are calculated by DFT using the Perdew-Burke-Ernzerhof (PBE) functional [34], which is implemented in the Quantum Espresso package [35]. The vdW interactions are included by the Grimme-D2 frame [36]. The plane-wave energy cutoff is set to be 80 Ry based on norm-conserving pseudopotentials with semi-core electrons included. We choose a 24×24×1 k-grid sampling in the reciprocal space. A vacuum distance is set to be around 18 Å between adjacent bilayers to avoid spurious interactions.

The quasiparticle energies and band gaps are calculated within the single-shot $G_0W_0$ approximation using the general plasmon pole model [21]. The optical absorption spectra and excitonic effects are calculated by solving the Bethe-Salpeter Equation (BSE) [22]. We use a k-point grid of 24x24x1 for calculating the *e-h* interaction kernel and a fine k-point grid of 96×96×1 for converged excitonic states and optical absorption spectra. These GW-BSE simulations are performed by using the BerkeleyGW code [37] including the slab Coulomb truncation. For optical absorption spectra, only the incident light polarized parallel along the plane is considered due to the depolarization effect along the out-of-plane direction [38]. The energy impact from the SOC is included by using the DFT corrections to the GW quasiparticle and exciton energies [19].

As shown in the schematic plots in Figures 1 (a) and (b), we consider two types of twisted $MoSe_2/WSe_2$ heterostructures. The *H* type derives from a small twist angle ($\theta$) rotated from the AA' stacking style that is the pristine structure of bulk $MoSe_2$ and $WSe_2$ [39,40]. The other *R* type represents a small twist angle ($\theta$) rotated from the AA stacking style that is essentially a 60° rotation from the AA' stacking structure. In these two types of twisted bilayers, six local stacking styles can be identified, as listed in Figures 1 (a) and (b). Following the notations of previous publications, [5,6] we denote them as $H_h^h$ (AA'), $H_h^X$, $H_h^M$, $R_h^h$ (AA), $R_h^M$, and $R_h^X$. The superscript symbols, $h, X, and\ M$, represent the hollow center of a hexagon, chalcogen, and transition-metal elements, respectively, of the upper layer ($MoSe_2$). The subscript symbol represents the atomic sites in the bottom layer. For example, $R_h^M$ represents a local site at which the transition metal of the upper layer ($MoSe_2$) is above the hollow center of the hexagon of the bottom $WSe_2$ layer.

**III. Quasiparticle energy and its variation**

A typical quasiparticle band structure calculated by the GW approximation [21] is presented in Figure 2 (a) for the $H_h^h$ stacking style. To focus on many-electron effects, SOC is not included at this stage, but it will be considered in section VI, where we draw quantitative conclusions and compare our results with measurements. We focus on the band gap at the K/K' point because the vertical interband transitions and excitons around these points are likely responsible for optical spectra observed from the small twist angle $MoSe_2/WSe_2$ heterostructures. [41] All these six local stacking structures exhibit a type-II band alignment. Particularly, enhanced quasiparticle self-energy corrections dominate the electronic structures. For example, because of the reduced screening in 2D structures, many-electron interactions significantly increase the 1.26 eV DFT band gap to be 1.93 eV for the $H_h^h$ stacking style.

Table I summarizes the GW-calculated quasiparticle band gaps at the K/K' point, which vary with the local stacking styles. For the *H* stacking styles, the energy variation is rather small: $H_h^X$ and $H_h^h$ has the smallest band gap around 1.93 eV, while $H_h^M$ has the largest band gap around 1.96 eV. The difference is more significant in the *R* twisting cases: $R_h^X$ has the smallest band gap around 1.89 eV while $R_h^M$ has the largest band gap around 1.99 eV, showing a 100-meV variation of the quasiparticle band gap. A larger band-gap variation is always preferred to confine free carriers for realizing those correlated physics. In this sense, our calculation indicates that correlation effects and moiré patterns of quasiparticles can be more significant in *R*-type $MoSe_2$/$WSe_2$ twisted bilayers, which are formed by a small twist angle from the AA stacking style. It must be pointed out that, because our simulation is based the interpolation approach [5] of regularly stacked bilayers, we cannot directly produce the model-predicted flat bands [7,11,42]. On the other hand, the calculated many-electron screening effect shall be similar for correcting energies of flat bands and useful for further constructing Hubbard models. [8]

### IV. Optical absorption spectra and excitons

Excitonic effects are known to be dramatically enhanced in 2D semiconductors, and they dictate the optical spectra. [19,43] Using the GW-calculated quasiparticle energy and dielectric function, we can solve the Bethe-Salpeter Equation (BSE) [22] to obtain excitonic states. The optical absorption spectra corresponding to these six local stacking styles are presented in Figures 2 b1-c3. Like many other 2D structures, enhanced excitonic effects are observed: after including *e-h* interactions, numerous excitonic peaks (red-color curves) are formed below the quasiparticle band gap with significant *e-h* binding energies around a few hundred meV.

Based on the spatial distribution of electron and hole wavefunctions, there are mainly two types of excitons, the intralayer and interlayer ones. First, we discuss intralayer excitons. In Figures 2 b1-c3, nearly all bright peaks, marked by $X_0$, are from intralayer excitons because of the significant overlap of their electron and hole wavefunctions and the corresponding large dipole oscillator strength. We also note that *e-h* binding energies of these intralayer excitons in twisted bilayers are smaller (around 100~200 meV) than those of monolayer structures. Take the $H_h^h$ stacking style as an example (Figure 2 (b1)). The lowest-energy intralayer exciton ($X_0$) is from the MoSe$_2$ layer, which is located at 1.65 eV. The quasiparticle band gap of this MoSe$_2$ layer in the heterostructure is around 2.08 eV. Thus, the *e-h* binding energy of this intralayer exciton is around 430 meV, which is smaller than that of suspended monolayer studied in previous works (0.65±0.1 eV). [16] This reduced intralayer *e-h* binding energy of the MoSe$_2$ layer is from the additional screening effect by the neighboring WSe$_2$ layer.

We now discuss interlayer excitons, which are the lower-energy excitations than intralayer excitons in MoSe$_2$/WSe$_2$ bilayers because of the type-II band alignment. In Figure 2, we mark the energy of the lowest-energy interlayer excitons ($IX_0$) by the black arrows. First, we focus on their energies. Interestingly, although the electron and hole wave functions are spatially separated into different layers, the *e-h* binding energy of interlayer excitons is not significantly smaller than that of intralayer excitons. Take the $H_h^h$ stacking structure as an example. The marked interlayer exciton ($X_0$) is located at 1.52 eV in Figure 2 (b1), and the quasiparticle band gap is around 1.93 eV, resulting in an *e-h* binding energy of 410 meV. This is close to that (430 meV) of the intralayer exciton of the MoSe$_2$ layer discussed above. These similar *e-h* binding energies of intralayer and interlayer excitons are due to the long-range nature of screened Coulombic interactions.

The optical oscillator strength of interlayer excitons is crucial for understanding experimental measurements. The optical oscillator strength of interlayer excitons is usually assumed to be small because of the separation of electron and hole wavefunctions due to the band offset. However, this assumption is not always true in twisted bilayers. For example, the marked interlayer excitons ($IX_0$) in Figures 2 (b1) and (c1) are visible and their dipole oscillator strengths are on the same order as those of intralayer excitons. On the other hand, the optical oscillator strengths of interlayer excitons can be dramatically changed by the stacking style. To make all these interlayer excitons visible, we plot their dipole oscillator strength in Figures 3 (a) and (b) on a logarithmic scale. The dipole oscillator strength can vary by five or six orders of magnitude within a supercell. The interlayer excitons formed in the $H_h^h$ and $R_h^h$ stacking styles are the brightest while the interlayer excitons of the $H_h^M$ and $R_h^M$ stacking styles are the darkest.

We further interpolate the local excitonic quantities calculated at the selected high symmetry points by the biharmonic spline method [44] for a smooth distribution. The generated moiré patterns of exciton dipole oscillator strength are presented in Figures 3 (c) and (d) for the $H$ and $R$ twisted bilayers, respectively. The dipole oscillator strengths form triangular lattices with a sharp contrast in both twisted structures. This huge variation of dipole oscillator strengths indicates that experimental measured optical signals, e.g., photoluminescence (PL), of interlayer excitons may be mainly decided by those located in the $H_h^h$ or $R_h^h$ area. Finally, these interlayer excitons inherit the valley-dependent selection rules [6], and the valley selection rules of specific interlayer excitons are marked in Figures 3 (a) and (b) as well.

The dipole oscillator strength is strongly correlated with the overlap between electron and hole wavefunctions. In Figures 4, we have plotted wavefunctions of two typical interlayer excitons: the bright one from the $H_h^h$ stacking style and the dark one from the $H_h^M$ stacking style. From the top view, these two excitons look very similar, exhibiting a 1$s$-like, spherical excitonic state with a size around 5 nm. However, their interlayer distributions are very different. For the dark interlayer exciton ($H_h^M$), electrons and holes are well separated into two layers. In Figure 4 (b2), when the electron is fixed on the top (MoSe$_2$) layer, the hole wavefunction is completely confined within the bottom (WSe$_2$) layer. As shown in Figure 4 (b3), when the hole is fixed in the bottom (WSe$_2$) layer, the electron wavefunction is completely confined within the top (MoSe$_2$) layer. Therefore, the overlap between electrons and holes is negligible, resulting in an optically dark exciton state. However, for the bright exciton located at the $H_h^h$ stacking style, a significant overlap between electron and hole wavefunctions is observed. In Figure 4 (a2), when the electron is fixed in the top (MoSe$_2$) layer, the hole wavefunction spreads over both layers. In Figure 4 (a3), when the hole is fixed in the bottom (WSe$_2$) layer, the electron wavefunction is completely confined within the upper (MoSe$_2$) layer. Therefore, the interlayer hybridization of valence (hole) bands results in the enhanced dipole oscillator strength of interlayer excitons in the $H_h^h$ stacking style.

## V. Radiative lifetime of interlayer excitons

The remarkable difference of oscillator strength of interlayer excitons between stacking styles lead to a significant variation of the intrinsic radiative lifetime of interlayer excitons. The radiative lifetime of an exciton in the $S$ state with a zero center-of-mass momentum at zero temperature can be derived from the Fermi's Golden rule as:

$$\tau_s^{-1}(0) = \frac{8\pi e^2 E_s(0) \mu_s^2}{\hbar^2 c A_{uc}} \quad (1),$$

where the $E_s(0)$ is energy of the exciton state $S$, $c$ is the speed of light, $\mu_s^2$ is the modulus square of the exciton dipole moment, and $A_{uc}$ is the area of the unit cell. [25,45]

To consider the effect of a finite temperature on the exciton radiative lifetime, one way is to average the decay rate ($\tau_s^{-1}$) over the momentum range thermally accessible at temperature $T$. Assuming that the exciton state $S$ has a parabolic dispersion $E_s(q) = E_s(0) + [(\hbar^2 q^2)/(2M_s)]$ with an exciton reduced mass $M_s$, the average radiative lifetime $<\tau_s>_T$ can be written as [45]

$$<\tau_s> = \tau_s(0) \frac{3}{4} \left(\frac{E_s(0)^2}{2M_s c^2}\right)^{-1} k_B T \quad (2)$$

Following these formulas, our calculated interlayer exciton radiative lifetimes at 77K, which is the temperature of liquid nitrogen, are also summarized in Table I. The radiative lifetime varies by more than seven orders of magnitude with different stacking styles in those $H$-type stacking structures and more than five orders of magnitude in those $R$-type stacking styles. This indicates that experimentally measured exciton lifetime may exhibit complicated origins. It has to be pointed out that phonons and other factors play a crucial role in limiting interlayer exciton lifetime at higher temperature, but calculations taken these factors into account are beyond the scope of our current work.

**VI. The corrections to the interlayer exciton energy and comparing with experiments.**

The above simulations do not include SOC and the minor lattice mismatch of the heterostructure although these factors will not change the fundamental physics pictures. On the other hand, we must consider them for quantitatively comparisons with measurements. First, we focus on the SOC corrections that reduce the band gap at the K/K' point. Here, we consider the SOC splittings from DFT simulations to correct the GW quasiparticle energies and excitons [19]. Take $H_h^h$ as an example, we present our quasiparticle band structures with SOC included from DFT corrections as well as lattice mismatch compensation in Figure 5 (a). The same color bar is used for representing the projected components from different layers. Actually, our calculation indicates that all these six stacking styles of MoSe$_2$/WSe$_2$ heterostructure still preserve a direct band gap at the K/K' point after considering SOC effect [41].

Second, we focus on the lattice-match corrections. The above simulations impose the same lattice constants on both layers of MoSe$_2$/WSe$_2$ heterostructures. As a result, compared with the calculated values of suspended monolayers, the MoSe$_2$ layer is slightly stretched by ~ 0.5% and the WSe$_2$ layer is slightly compressed by ~ 0.7%. This will impact the band alignment slightly. To eliminate this artificial lattice match, we first calculate the absolute energy of CBM and VBM of each relaxed monolayer according to the vacuum level. Then we add the same strain (in bilayer calculations) to each layer and find the energy shifts. With these energy shifts, we can modify the band alignment and correct the energies of quasiparticle band gaps and optical spectra. As shown in Figure 5 (b), the DFT-calculated absolute energies of CBM and VBM of a relaxed MoSe$_2$ monolayer is about -3.88 eV and -5.21 eV, respectively. After considering the 0.5% stretch to match the lattice constant of the heterostructure, we find that the CBM and VBM are shifted to be -3.95 eV and -5.23 eV, respectively. Similarly, we can obtain the energy shifts for the WSe$_2$ layer. The modified band alignment is shown in Figure 5 (b). The DFT band gap ($E_g$) is reduced by 0.24 eV (from 1.02 eV to 0.88 eV) because of this artificial lattice matching in our calculations ($E_{strained}$). Given that these energy corrections to band edges are higher-order corrections, [46] we can increase energies of quasiparticle interlayer band gap and interlayer exciton by this correction, 0.24 eV to eliminate the influence of the artificial lattice match introduced in our simulations.

With both effects included, the finalized DFT, quasiparticle valence band maximum (VBM) referring to the vacuum level at the Γ and K points, quasiparticle band gaps, and exciton energies are summarized in Table II. [47] These VBM values can be useful for further constructing corresponding Hubbard model and study the twisted-angle-induced flat bands. [8] Importantly, the calculated energies of interlayer exciton are in excellent agreements with recent measurements of exciton resonances observed in MoSe$_2$/WSe$_2$ bilayers [11,12,15,41]. For example, our calculated lowest-energy interlayer exciton for the *R*-type twisted MoSe$_2$/WSe$_2$ bilayer is around 1.36 eV, which is very close to recent photoluminescence (PL) measurements, 1.31~1.40 eV [11,12,15].

Finally, we summarize the energies of interlayer excitons with all above factors included. As shown in Figures 6 (a) and (b), the interlayer exciton in the $H_h^h$ region has the lowest energy (1.41 eV) in *H*-type twisted structures, while the interlayer exciton in the $R_h^X$ region has the lowest energy (1.36 eV) in *R*-type twisted structures. The corresponding moiré patterns of interlayer exciton energy are presented in Figure 6 (c) and (d). More specifically, the *H*-type twisted bilayers exhibit a weaker confinement of excitons: the low-energy blue area is dominant, and the confinement height is around 30 meV. The *R*-type twisted structures exhibits a stronger confinement: The low-energy blue area is isolated by higher energy barriers, and the confinement height is around 100 meV. Therefore, the *R*-type twisted structure is preferred for realizing quantum-dot arrays and observing lateral quantum confinement effects of interlayer excitons.

## VII. Conclusion

In conclusion, we investigate quasiparticle energies and excitons of MoSe$_2$/WSe$_2$ bilayers with a small twist angle using the first-principles MBPT. The quasiparticle energy and interlayer exciton energy vary by up to 100 meV between different local sites within a moiré supercell, and this energy modulation is more significant in *R*-type twisted structures rotated slightly from the AA stacking style. Despite the nearly spatially homogeneous *e-h* binding energies, our calculations reveal that different local stacking styles can dramatically modify the dipole oscillator strength and radiative lifetime of interlayer excitons by a few orders of magnitude. As a result, optical moiré patterns with high contrast are predicted in both *H* and *R* type twisted structures. Our results are helpful for interpreting optical spectroscopy measurements performed in the far-field and guide future near-field experiments to identify moiré-potential confined optical excitations. Furthermore, these predictions provide quantitative guidance to realize the proposed many-body physics and to search for quantum emitters in twisted vdW bilayers.

**Acknowledgement**

The work at Washington University are supported by the National Science Foundation (NSF) CAREER Grant No. DMR-1455346 and the Air Force Office of Scientific Research (AFOSR) grant No. FA9550-17-1-0304. The computational resources have been provided by the Stampede of TeraGrid at the Texas Advanced Computing Center (TACC) through XSEDE. X. Li gratefully acknowledges support from NSF DMR-1808042 and the Welch Foundation via grant F-1662.

**Tables:**

Table I: Band gaps, excitons, and radiative lifetimes of MoSe$_2$/WSe$_2$ heterostructures without including SOC and lattice mismatch.

| Stacking styles | DFT band gap at K/K' point (eV) | Quasiparticle band gap at K/K' point (eV) | Interlayer exciton $IX$ (eV) | Radiative lifetime at 77K (S) |
|---|---|---|---|---|
| $H_h^h$ | 1.26 | 1.93 | 1.52 | $2.5 \times 10^{-9}$ |
| $H_h^X$ | 1.26 | 1.93 | 1.54 | $1.1 \times 10^{-6}$ |
| $H_h^M$ | 1.26 | 1.96 | 1.58 | $5.9 \times 10^{-2}$ |
| $R_h^h$ | 1.26 | 1.96 | 1.57 | $3.5 \times 10^{-9}$ |
| $R_h^M$ | 1.32 | 1.99 | 1.59 | $8.8 \times 10^{-4}$ |
| $R_h^X$ | 1.22 | 1.89 | 1.50 | $2.2 \times 10^{-7}$ |

Table II: Band gaps, quasiparticle VBM (referring to vacuum), and interlayer excitons after including SOC and lattice mismatch. The last column are data from measurements.

| Stacking styles | DFT band gap at K/K' point (eV) | VBM (eV) $\Gamma$ | VBM (eV) K/K' | Quasiparticle band gap at K point (eV) | Interlayer exciton $IX$ (eV) | Experimental PL interlayer exciton (eV) |
|---|---|---|---|---|---|---|
| $H_h^h$ | 1.14 | -5.50 | -5.43 | 1.82 | 1.41 | 1.32~1.40 [12,41] |
| $H_h^X$ | 1.13 | -5.49 | -5.40 | 1.80 | 1.41 | |
| $H_h^M$ | 1.11 | -5.70 | -5.43 | 1.82 | 1.44 | |
| $R_h^h$ | 1.12 | -5.70 | -5.41 | 1.82 | 1.43 | 1.31~1.40 [11,12,15] |
| $R_h^M$ | 1.19 | -5.50 | -5.46 | 1.86 | 1.46 | |
| $R_h^X$ | 1.08 | -5.48 | -5.40 | 1.75 | 1.36 | |

**Figures**

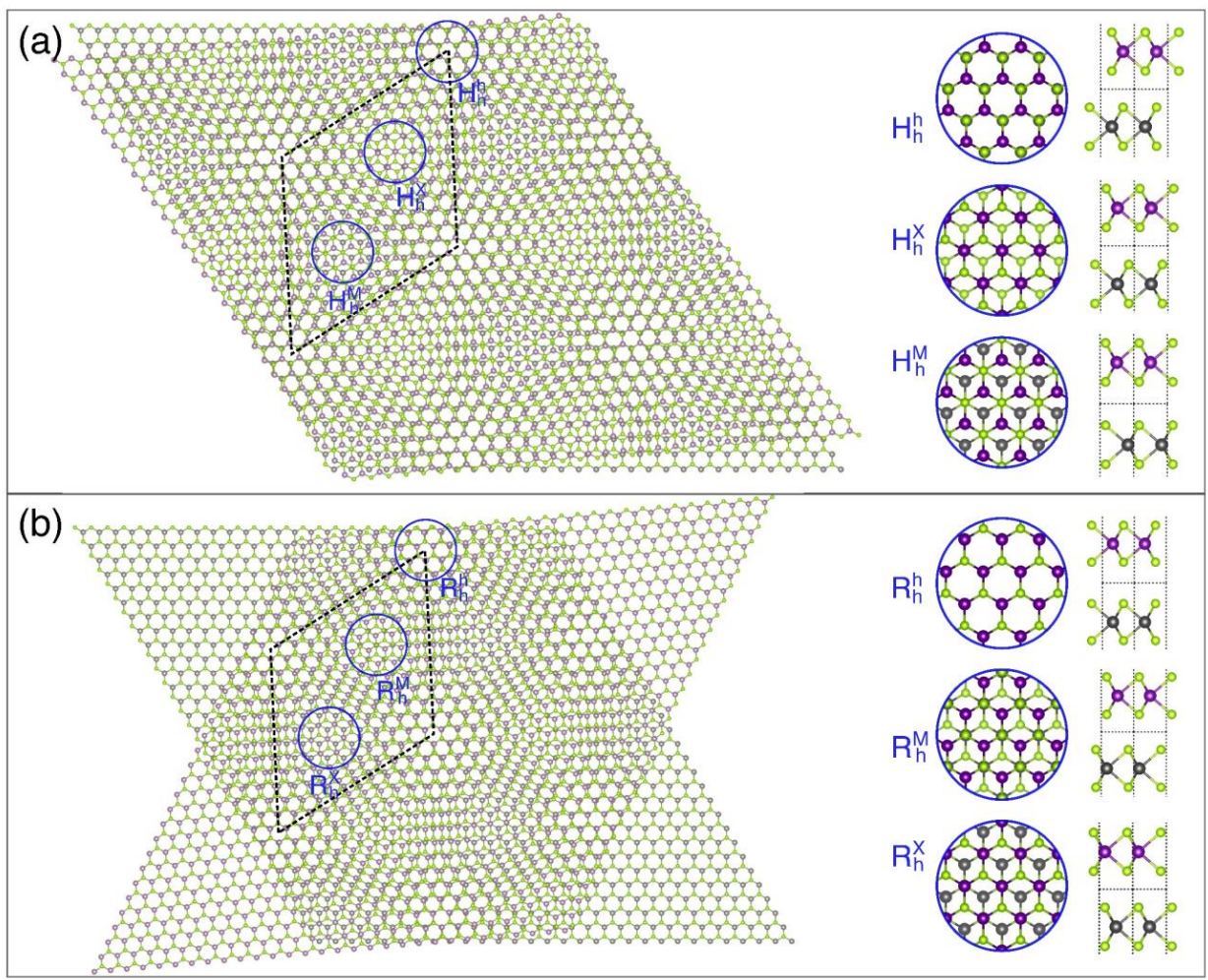

**Figure 1** The schematic plots of two main types of twisting MoSe$_2$/WSe$_2$ bilayer heterostructures. (a) is the *H*-type twisting structure rotated from the AA' stacking style, and (b) is the *R*-type twisting structure rotated from the AA stacking style. Six local stacking styles are identified and amplified with top and side views.

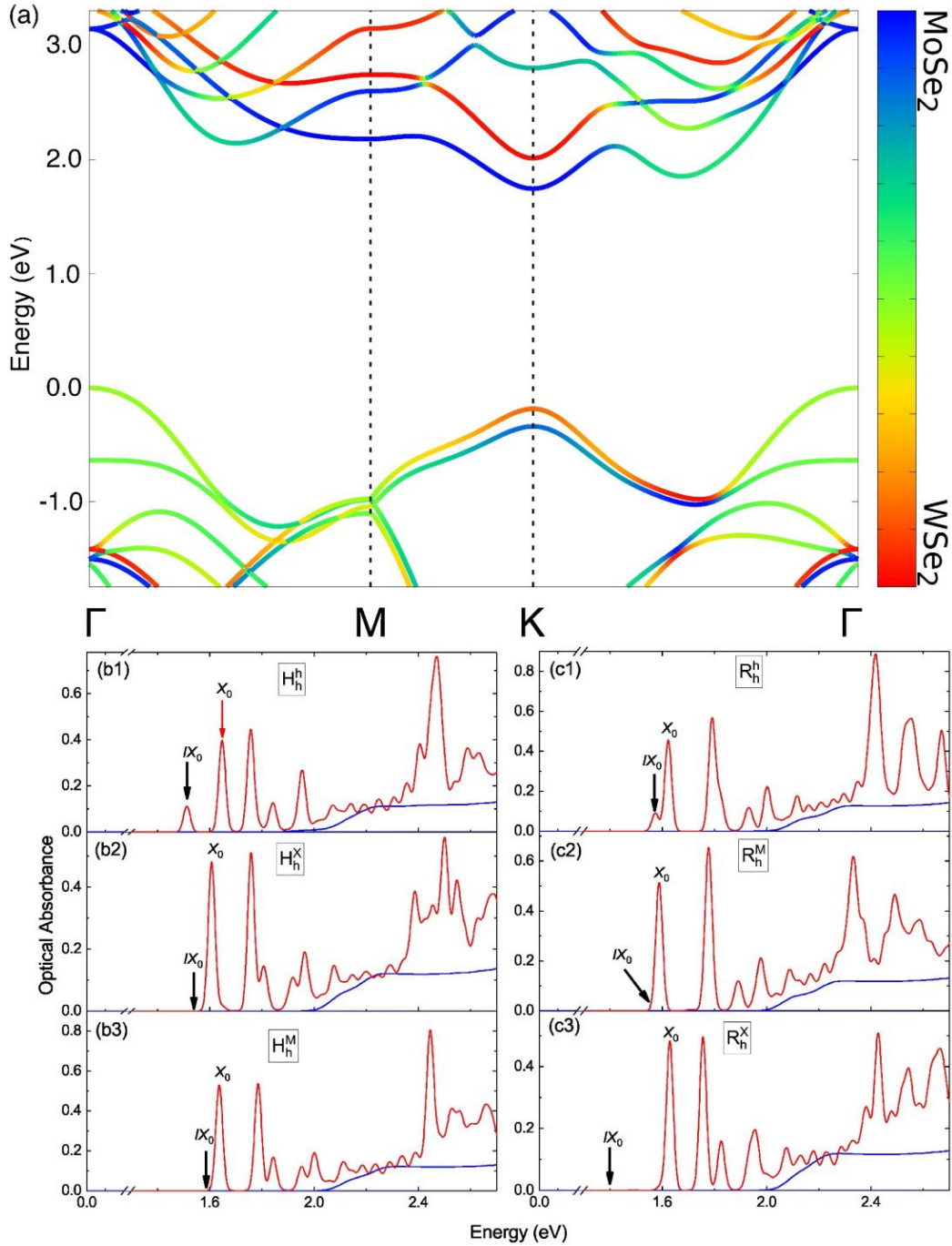

**Figure 2** (a) Quasiparticle band structure of the $H_h^h$ twisted MoSe$_2$/WSe$_2$ bilayer. The energy of the top of valence band is set to be zero. The projected components of electronic states to those of MoSe$_2$ and WSe$_2$ are represented by different colors, respectively. (b1-b3 and c1-c3) Optical absorption spectra (with (red) and without *e-h* (blue) interactions) of six identified local stacking styles in a MoSe$_2$/WSe$_2$ heterostructure. The energy of the interlayer exciton ($IX_0$) is marked by the black arrow. SOC is not included. A 13-meV smearing to spectral widths is applied.

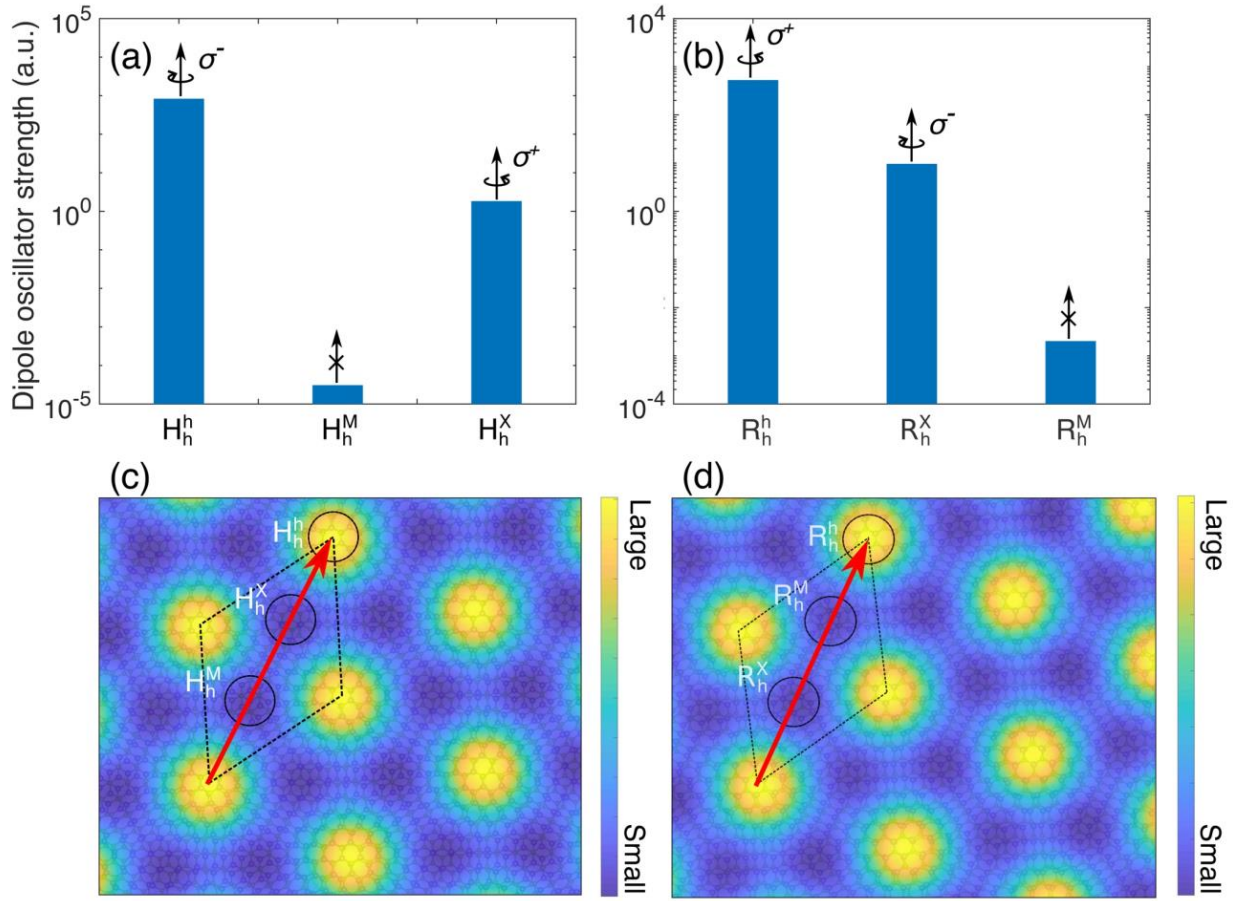

**Figure 3** (a) and (b) are optical oscillator strength of interlayer excitons at the six local sites with labeled stacking styles displayed on the logarithmic scale, respectively. (c) and (d) are the interpolated dipole oscillator strength of the *H*-type and *R*-type twisted MoSe$_2$/WSe$_2$ bilayers, respectively. The valley-dependent selection rules of interlayer excitons are marked in (a) and (b) according to Ref [6].

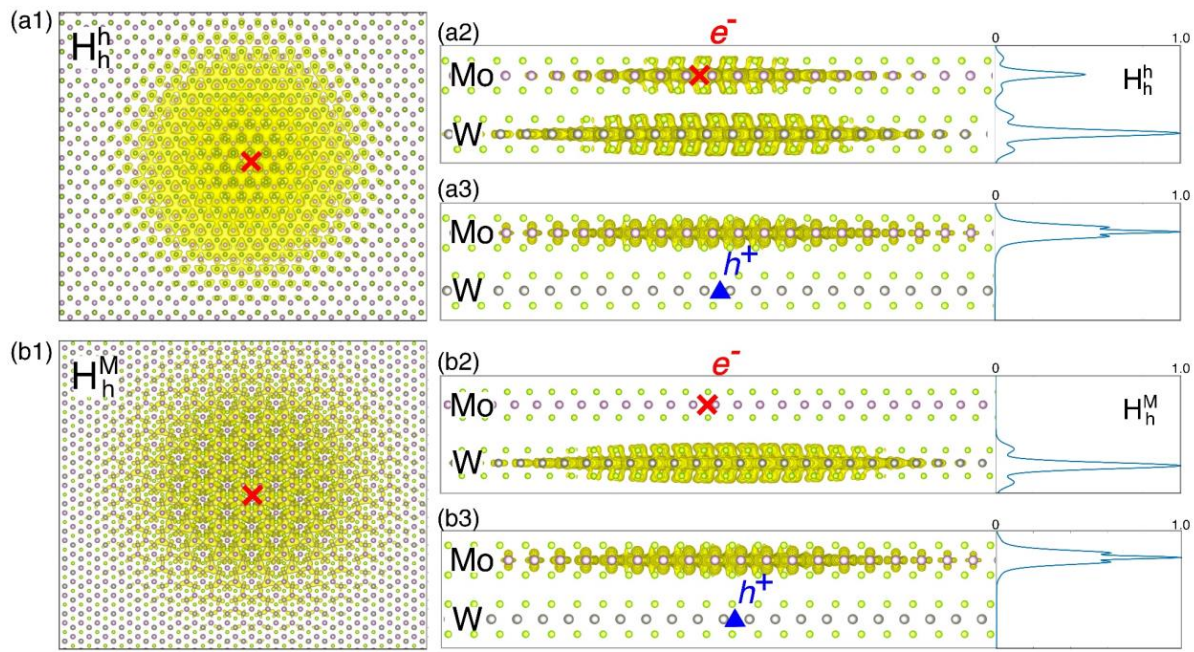

**Figure 4** Interlayer exciton wavefunctions of the $H_h^h$ and $H_h^M$ stacking styles. (a1) and (b1) are top views of the exciton wavefunctions. (a2) is the side view of the bright interlayer exciton. The electron (the cross) is fixed in the upper (MoSe$_2$) layer, and the hole wavefunction is plotted. The side panel is the integrated hole wavefunction. (a3) is the same exciton of (a2), but we fix the hole (the triangle) in the lower (WSe$_2$) layer and plot the electron wavefunction. The side panel is the integrated electron wavefunction. (b2) and (b3) are the similar plots for the dark interlayer exciton at the $H_h^M$ stacking style.

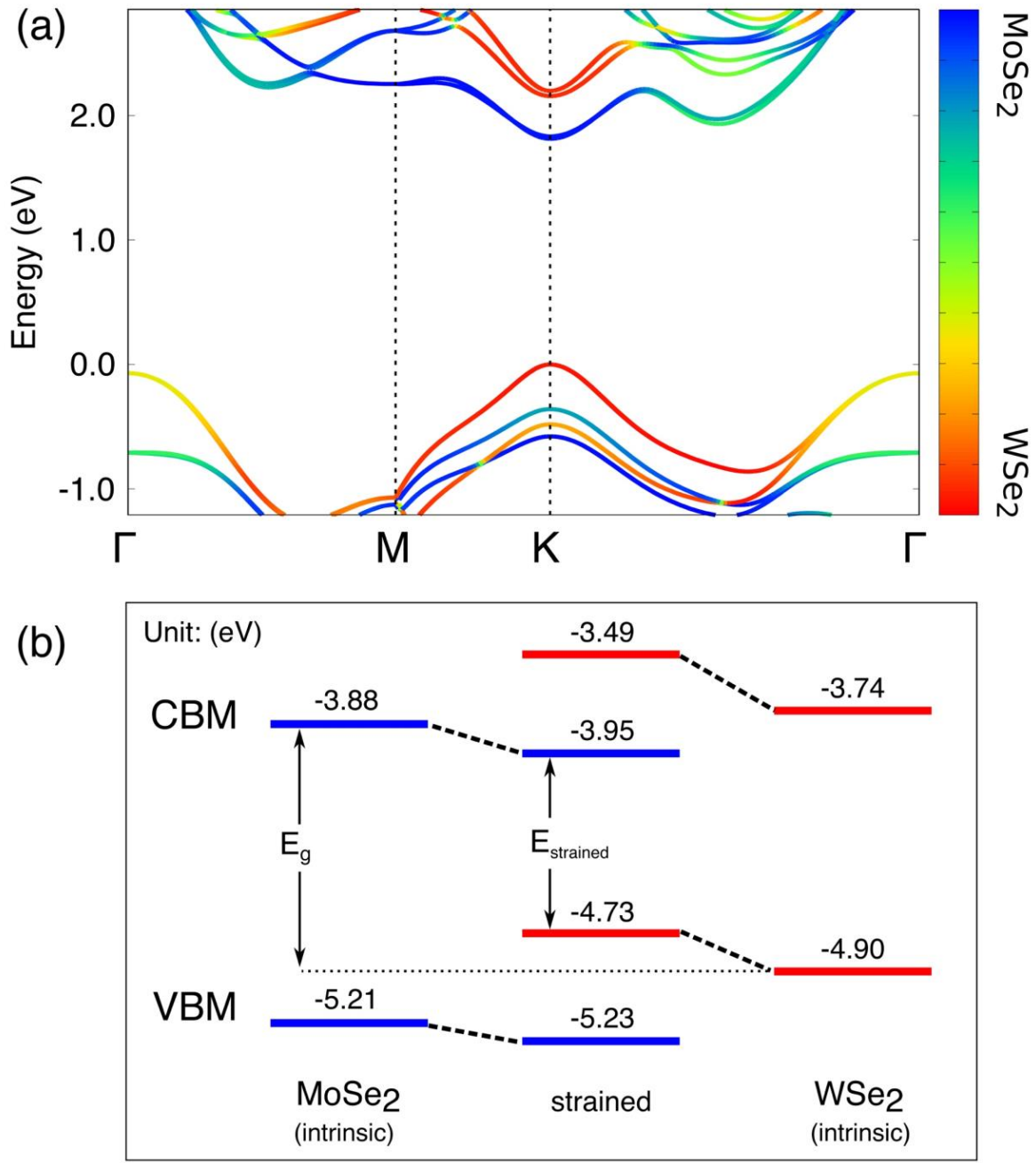

**Figure 5** (a) Quasiparticle band structure of the $H_h^h$ twisted bilayer MoSe$_2$/WSe$_2$ heterostructure with SOC included. The energy of the top of valence band is set to be zero. The projected components to the MoSe$_2$ and WSe$_2$ are represented by different colors. (b) Schematic band alignment of the CBM and VBM for unstrained monolayer (MoSe$_2$ and WSe$_2$) and strained ones to match the lattice constant of heterostructures. The absolute band energies are DFT results.

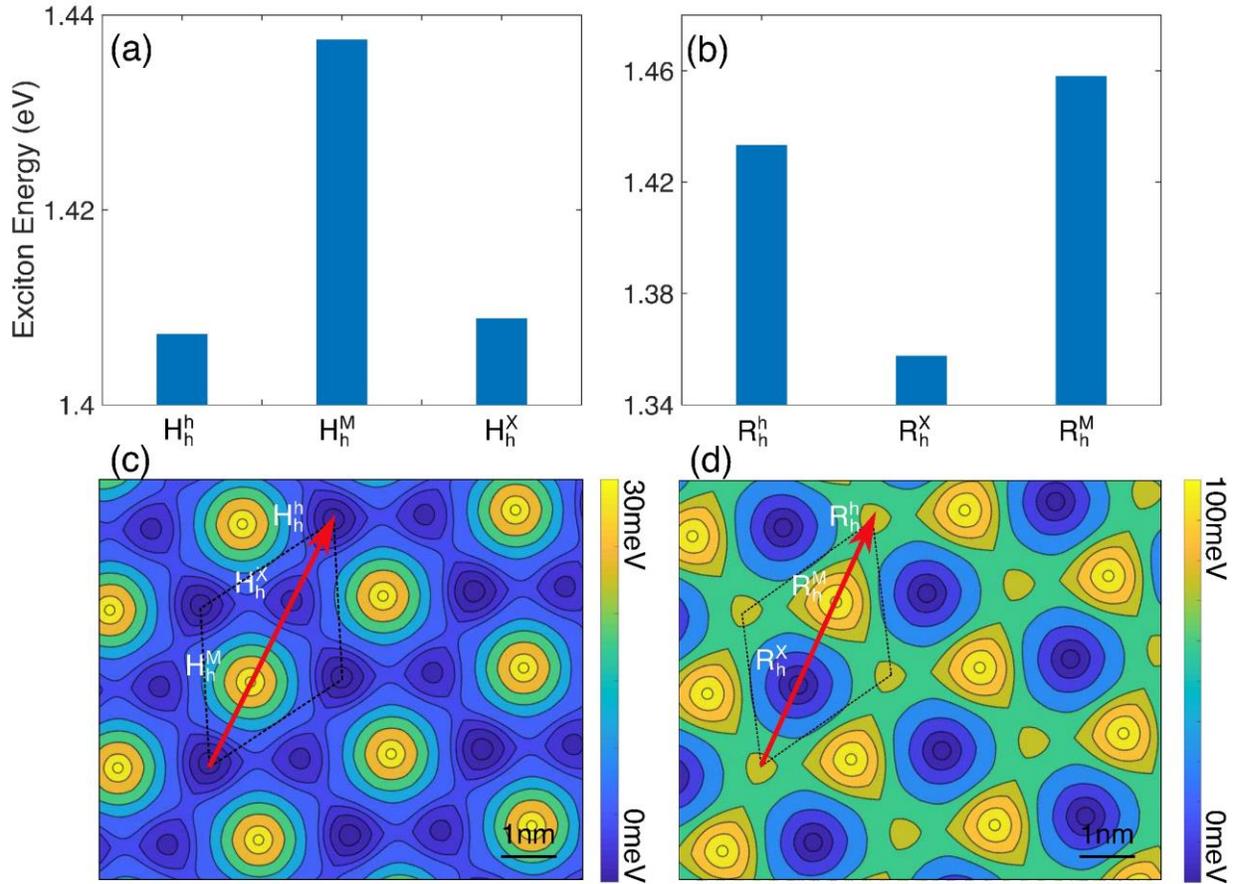

**Figure 6** (a) and (b) are energies of interlayer excitons at the six local sites *H*-type and *R*-type twisted MoSe$_2$/WSe$_2$ bilayers, respectively. (c) and (d) are the interpolated interlayer exciton energy of the *H*-type and *R*-type twisted bilayers, respectively. The lattice mismatch and SOC are included.